\documentclass[lettersize,journal]{IEEEtran}
\usepackage{amsmath,amsfonts}
\usepackage{algorithmic}
\usepackage{algorithm}
\usepackage{array}
\usepackage[caption=false,font=normalsize,labelfont=rm,textfont=rm]{subfig}
\usepackage{textcomp}
\usepackage{stfloats}
\usepackage{url}
\usepackage{verbatim}
\usepackage{graphicx}
\usepackage{cite}
\usepackage{amsmath}
\usepackage{amsthm} 
\usepackage{multirow}

\usepackage{enumitem} 
\usepackage{xcolor}  
\usepackage{soul}    
\setulcolor{blue}
\theoremstyle{definition}
\usepackage{mathrsfs}
\usepackage{pifont}

\usepackage{listings}
\usepackage{xcolor}
\usepackage[table]{xcolor}  
\usepackage{array}

\usepackage{caption}
\DeclareCaptionFont{redlabel}{\color{red}}

\begin{document}

\title{ Channel Ordering for Fairness in Elastic Optical Networks via a LLM-Guided Bottleneck TSP Solver}

\author{Liangshun Wu,~\IEEEmembership{Member,~IEEE,} Wen Chen*,~\IEEEmembership{Senior Member,~IEEE,} Qingqing Wu,~\IEEEmembership{Senior Member,~IEEE,}
\thanks{Liangshun Wu, Wen Chen and Qingqing Wu are with the School of  Information and Electronic Engineering, Shanghai Jiao Tong University, Shanghai 200240, China (Email: \{wuliangshun,wenchen,qingqingwu\}@sjtu.edu.cn).}
\thanks{* Wen Chen is the corresponding author.}
\thanks{This work is supported by NSFC 62531015 and U25A20399, and Shanghai Kewei 24DP1500500.}}



\maketitle

\begin{abstract}
In flexible-grid elastic optical networks (EONs), the ordering of frequency channels plays a crucial role in managing inter-channel interference and ensuring signal quality. We address the Channel Ordering Problem (COP) by reformulating it as a Bottleneck Traveling Salesman Problem (BTSP), where interference among channels is represented as edge weights in a graph structure. To tackle this challenge efficiently, we develop a scalable approach that integrates statistical exploration with guidance from large language models (LLMs). Extensive simulations using both the Gaussian Noise (GN) model and the split-step Fourier method demonstrate that our method achieves near-optimal signal-to-noise ratio (SNR) performance and offers robust scalability across diverse network settings, making it well-suited for practical deployment in large-scale optical communication systems.
\end{abstract}


\section{Introduction}
\IEEEPARstart{I}{n}  flexible-grid elastic optical networks (EONs), channel placement within the frequency spectrum significantly affects inter-channel interference and overall signal quality. As described in \cite{bhar2019channel}, the channel ordering problem (COP) involves finding a permutation of the available channels on the frequency grid that maximizes the minimum signal-to-noise ratio (SNR) across all channels, thereby maximizing the network’s SNR margin. This is essential for meeting quality-of-transmission requirements, but the problem is highly complex: the number of possible orderings grows factorially with the channel count, making exhaustive search infeasible for realistic networks. In fact, COP is NP-hard, necessitating more tractable approaches than brute-force enumeration. To address this challenge, we reformulate COP as a graph problem: each channel is represented as a vertex in a complete graph, and edges represent the SNR penalty (noise increase) between two adjacent channels on the frequency axis. Maximizing the minimum SNR then becomes equivalent to finding a Hamiltonian path that minimizes the maximum edge weight — in other words, a bottleneck Traveling Salesman Problem (BTSP) on the channel graph.

The TSP is a classic NP-hard problem. Exact solvers (e.g., cutting-plane, dynamic programming \cite{Qin1994Bellman}, branch-and-bound \cite{kinable2017exact}) guarantee optimality but are computationally prohibitive for large instances \cite{applegate2006traveling}. To scale up, numerous heuristics and meta-heuristics—such as Lin–Kernighan \cite{helsgaun2000effective}, BLKH \cite{HelsgaunBTSP}, Larusic-Punnen\cite{Larusic2012}, ant colony optimization, simulated annealing, genetic algorithms, and neural networks—have been developed to yield near-optimal solutions, albeit without guarantees. Recently, Large Language Models (LLMs) have been explored for combinatorial optimization via prompt engineering and chain-of-thought (CoT) reasoning. Methods like OPRO \cite{yang2023large} generate direct TSP solutions, while FunSearch \cite{romera2024mathematical} and EoH \cite{liu2024evolution} evolve heuristic algorithms instead, yielding stronger generalization. Despite this progress, LLMs still lag on large instances due to limited feedback, a gap addressed by ReEvo \cite{ye2024reevo} through a feedback-driven evolutionary framework.


{Existing methods are not well suited to our setting}: generic heuristics are not designed for the bottleneck objective (e.g., GA crossover/mutation lack min–max rationale); 2-opt–style local search {often} stalls in local minima; RL-based solvers waste {computational} resources; {and while LLMs are conveniently accessible via} remote APIs, existing LLM pipelines (e.g., FunSearch\cite{romera2024mathematical}) are {overly} complex. In contrast, we propose a lightweight hybrid that {combines} bottleneck-aware probabilistic sampling with simple LLM-seeded heuristics to efficiently achieve near-optimal channel ordering for minimizing worst adjacent interference in EONs.

\section{ Channel Ordering as a Bottleneck TSP}

\begin{figure*}[h]
\centering
\includegraphics[width=\linewidth]{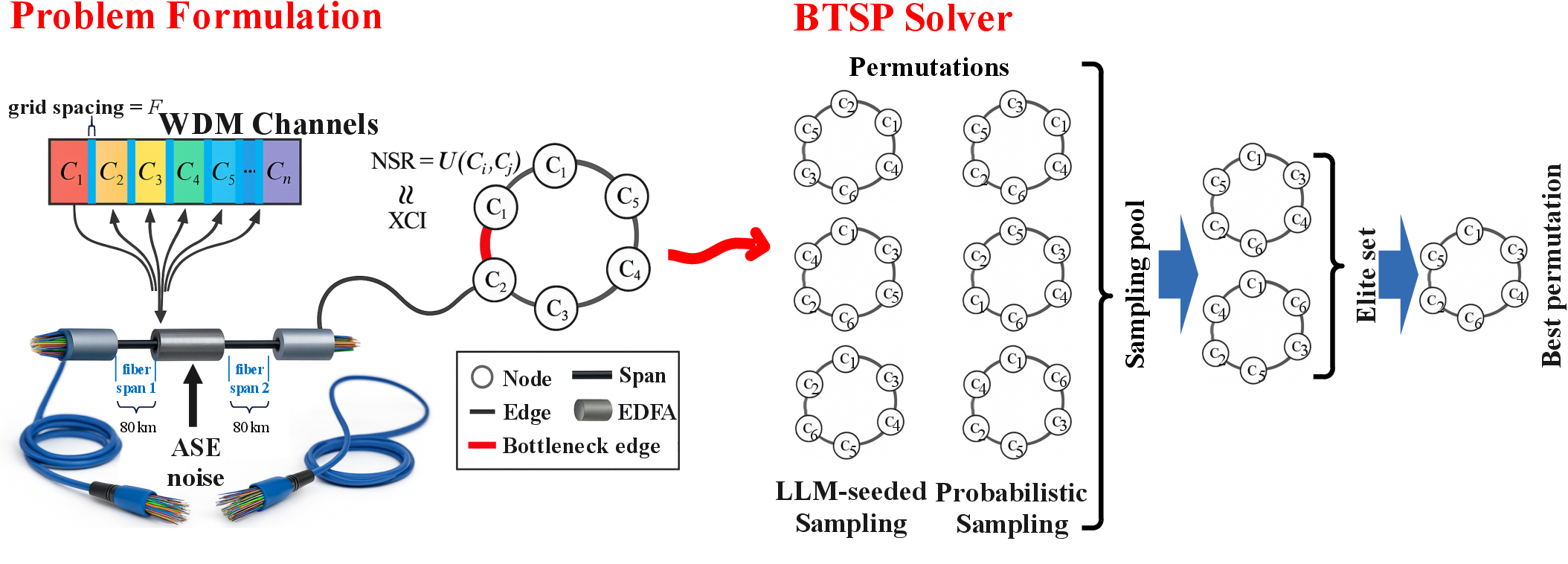}
\caption {
Left: Problem formulation: a fiber span chain with Erbium-Doped Fiber Amplifiers (EDFAs) introducing ASE noise; WDM channels $C_1,\dots,C_n$ with fixed grid spacing $F$ are mapped to a graph with symmetric XCI $U(C_i,C_j)$ as edge. The task is to find a Hamiltonian cycle/path that minimizes the maximum adjacent NSR (bottleneck edge highlighted). Right: The proposed BTSP solver fuses node-wise probabilistic sampling with LLM-{seeded} heuristics, and elite retention for statistics-guided probability refinement, {obtaining final best permutation}.}
\label{fig:sys}
\end{figure*}

\subsection{System Model}

Consider $n$ coherent {wavelength division multiplex} (WDM) channels $\mathcal{C}=\{C_1,\ldots,C_n\}$ on a single link with fixed grid spacing $F$, propagating over $N_s$ identical fiber spans (shown as Fig. \ref{fig:sys}). Each channel $C_i$ is characterized by its power $p(C_i)$, bandwidth $\Delta(C_i)$, and resulting power spectral density (PSD) $G(C_i) = p(C_i)/\Delta(C_i)$. 

Prior work emphasizes that the Enhanced Gaussian Noise (EGN) corrections, which {add} the modulation-dependent {higher-order field-moment terms} to the Gaussian Noise (GN) model, {explicitly handle} format dependence {(see \cite{Poggiolini, roberts2018measurement})}. Using the {EGN} model, the per-span cross-channel nonlinear interference (XCI) PSD generated on channel {$C_{\pi_i}$} (given an permutation {$\mathbf{P}=[C_{\pi_1},\ldots,C_{\pi_n}]$} of the channels) is expressed as

{
\begin{equation}
\begin{aligned}
G_{\mathrm{sp}}^{\mathrm{NLI}}({C_{\pi_i}}) &= \frac{3 \gamma^2 {\Phi_{\pi_i}} G({C_{\pi_i}})}{2\pi\alpha|\beta_2|} 
\Bigg[{\Phi_{\pi_i}}G^2({C_{\pi_i}})  \sinh^{-1} \\
& \Big|\frac{\pi^2 [\Delta({C_{\pi_i}})]^2 \beta_2}{\alpha}\Big|
 + \sum_{\substack{j=1\\j\neq i}}^n {\Phi_{{\pi_i},{\pi_j}}}G^2({C_{\pi_j}})\\
 & \ln\frac{|{\pi_i-\pi_j}|F + \Delta({C_{\pi_j}})/2}{|{\pi_i-\pi_j}|F - \Delta({C_{\pi_j}})/2}\Bigg],
\end{aligned}
\label{eq:NSR}
\end{equation}
}
{Here $\Phi_{\pi_i}$ (and $\Phi_{{\pi_i},{\pi_j}}$) depend on the modulation of channel $C_{\pi_i}$ (or channel $C_{\pi_j}$). For common formats, $\Phi=1$ for PM-QPSK, $\Phi\approx17/25$ for PM-16QAM, etc. } $\gamma$ is the fiber nonlinear coefficient, $\alpha$ is the attenuation, and $\beta_2$ is the group-velocity dispersion. Accumulating nonlinear interference (NLI) over $N_s$ spans and adding amplifier spontaneous-emission (ASE) noise with PSD $G^{\mathrm{ASE}} = (e^{\alpha L}-1)h\nu n_{\mathrm{sp}}N_s$,  {where $h$ is Planck constant, $L$  is the length of a span, $n_{sp}$ denotes the spontaneous emission factor of the amplifier, $\nu$ denotes the frequency of light}. The SNR for ${C_{\pi_i}}$ is {\cite{fang2024reconfigurable}}:
\begin{equation}
\mathrm{SNR}({C_{\pi_i}}) = \frac{G({C_{\pi_i}})}{G^{\mathrm{ASE}} + N_sG_{\mathrm{sp}}^{\mathrm{NLI}}({C_{\pi_i}}),} {\times 10^{- t_p/10}}.
\label{eq:GN}
\end{equation}
{where $t_p$ denotes transceiver penalty}. We define the noise-to-signal ratio (NSR) of ${C_{\pi_i}}$ as $\mathrm{NSR}({C_{\pi_i}}) = 1/\mathrm{SNR}({C_{\pi_i}})$. The worst-case SNR of a channel permutation is
\begin{equation}
{\widetilde {\mathrm{SNR}}} = \min_i \mathrm{SNR}({C_{\pi_i}}),
\end{equation}
equivalently,
\begin{equation}
{\widetilde {\mathrm{NSR}}} = \max_i \mathrm{NSR}({C_{\pi_i}}).
\end{equation}
Our objective is to minimize ${\widetilde {\mathrm{NSR}}}$ (or maximize ${\widetilde {\mathrm{SNR}}}$) by finding an optimal permutation $\mathbf{P}$ of the channels.

\subsection{ Graph Formulation}
{Assuming only adjacent channels interfere}, we can model the COP on a directed complete graph $\vec{\mathcal{G}}$ whose vertices correspond to channels, and the directed edge weight is XCI. In principle, the COP then seeks a Hamiltonian path through $\vec{\mathcal{G}}$ that minimizes the maximum edge weight (i.e., minimizes the worst-case adjacent-channel NSR).  To obtain a tractable approximation, we consider an undirected relaxation. Define a symmetric interference metric between any two channels:
\begin{equation}
U(C_{\pi_i},C_{\pi_j}) = \max\{\mathrm{NSR}(C_{\pi_i},C_{\pi_j}), \mathrm{NSR}(C_{\pi_j},C_{\pi_i})\},
\end{equation}
this yields an undirected complete graph $\bar{\mathcal{G}}$. Now, consider any Hamiltonian cycle $\mathbf{P}$ on $\bar{\mathcal{G}}$, the COP is thus approximated by the following BTSP formulation:
\begin{equation}
\min_{\text{Hamiltonian cycle }\mathbf{P}\subset\bar{\mathcal{G}}} \max_{i} U \big({C_{\pi_i}}, {C_{(\pi_i \bmod n)+1}}\big). 
\end{equation}

In words, we seek a cyclic ordering of the channels that minimizes the maximum pairwise interference between adjacent channels. This Hamiltonian cycle can then be converted to a Hamiltonian path solution for the original COP (by breaking the cycle at the point of maximum $U$).  {Please note that we study TSP solver rather than convex combinational optimization; thus a direct benchmark with convex optimization approaches (e.g.,  \cite{roberts2016convex}) is out of scope.}

\section{The Proposed BTSP Solver}
\subsection{ Algorithm Overview}
We solve the above BTSP with a combination of probabilistic search and LLM-generated heuristics. It operates in two phases: Probabilistic Sampling, and Refinement.

\begin{itemize}
 \item Probabilistic Sampling Phase: 
At each step, for the current channel $C_i$, the next channel $C_j \in \Omega_i$ is selected based on its expected contribution to the path’s bottleneck cost. Specifically, we define a risk-aware score:
\begin{equation}
s_j = \max \left\{ U(C_i, C_j), \max_{C_k \in \Theta_j} U(C_j, C_k) \right\},
\end{equation}
which estimates the potential maximum edge weight incurred by choosing $C_j$ next, where $\Theta_j$ is the remaining available channels of $C_j$. The selection probability is then given by:
\begin{equation}
p_j = \frac{\exp(-\mu  s_j)}{\sum_{C_\ell \in \Theta_i} \exp(-\mu  s_\ell)}.
\end{equation}
Here, $\mu$ is a temperature parameter that decays from $\mu_0$ to $\mu_{\min}$ by factor $\sigma$, {$\mu \gets \max\{\mu_0 - \sigma t,\, \mu_{\min}\}$}, gradually shifting sampling from between exploration {(with smaller $\mu$, the selection is more random)} to exploitation {(with larger $\mu$, the algorithm favors the currently best edges)}. After generating $n_{\mathrm{samp}}$ sequences, the top $n_{\mathrm{keep}}$ with the lowest bottleneck cost are retained {as ``elite set''}.

\item Refinement Phase: We refine the search by generating another $n_{\mathrm{samp}}$ sequences, now biased by historical information from the previous phase. We adjust the edge-selection probabilities using the past occurrence counts of edges. Specifically, if $\delta_{i j}$ is the number of times channel $j$ followed channel $i$ in the retained sequences, we weight the selection probability for $j$ by $\delta_{ij}$:
\begin{equation}
  p_j^{(\mathrm{wtd})} = \frac{\delta_{ij} p_j}{\sum_{\ell \in \Theta_i} \delta_{i\ell}\, p_\ell}\,. 
\end{equation}

We progressively reduce the number of sequences kept from $n_{\mathrm{keep_0}}$ by a factor $z$ in each iteration to a minimum of $n_{\mathrm{keep}_\text{min}}$, {i.e., $n_{\mathrm{keep}} \gets \max\{n_{\mathrm{keep_0}} - z t, n_{\mathrm{keep_{\min}}}\}$}, focusing the search on the most promising candidates. {This gradually shrinks the candidate pool and accelerates convergence by intensifying the search around the best solutions found.} Finally, for each candidate sequence we perform an exhaustive local search on the last $f$ channels (i.e., try all permutations of the last $f$ channels{, with $f\in \{2,3,4\}$}) to fine-tune the arrangement and minimize the bottleneck edge weight.  {Here, default: $n_{\mathrm{samp}}=3n^2$, $n_{\mathrm{keep}}=3n$}.
\end{itemize}

\subsection{ LLM-Guided Sampling}
As shown in Fig. \ref{fig:sys} (right), we query an LLM with a structured prompt that includes the channel list, the pairwise interference matrix, a plain-language statement of the goal—minimize the largest adjacent interference—and an output schema requesting several short runnable programs:\begin{quote}\small
\texttt{Given a channel list and a pairwise interference matrix : \texttt{\textbf{{{[example graph]}}}}, propose channel orderings that minimize the largest adjacent interference in a cycle. Return 5 short programs, each under 30 lines. }
\end{quote} 
We use DeepSeek-R1 API with decoding paramters: temperature = 0.2, top-p = 0.95, max-tokens = 512, single sample per prompt.
It returns compact heuristics such as \emph{MinMax-Greedy}, \emph{Beam-MinMax}, and \emph{Cluster-Bridge}. Each program produces a few {permutations}, which we score by their bottleneck interference; the strongest candidates are injected into the sampling pool. This seeding accelerates convergence.  LLM calls are one-shot per instance, so this does not change the solver’s complexity.

\section{Simulation}

\subsection{ Simulation Setup}
We use both the GN model and exhaustive split-step Fourier method (SSFM) for verification. Each WDM channel carries 10,000 QAM symbols in the simulation. The fiber link consists of five 80-km spans {($N_s=5$, $L=80$)}, with channel spacings $F \in \{50,100,200\}$ GHz. For simplicity, each channel’s bandwidth $\Delta(C_i)$ is set equal to a fixed fraction of $F$ (independent of the number of channels, i.e., we assume the baud rate and spectral shape remain constant). The fiber parameters are:  average power $\bar{p}$=0  dBm, nonlinear coefficient $\gamma = 1.32\times10^{-3}$ (W·m)$^{-1}$, attenuation $\alpha = 0.22$ dB/km, and group-velocity dispersion $\beta_{2} = -21.7$ ps$^2$/km. The optical carrier frequency is $\nu = 193.55$ THz and Planck’s constant $h = 6.63\times10^{-34}$ J·s. Each amplifier has a noise figure of 3.16 dB (spontaneous emission factor $n_{\mathrm{sp}} = 1.58$). We include a transceiver penalty of $t_{p}=2.5$ dB (this also accounts for polarization-mode dispersion under a high-duty-cycle Nyquist pulse shape, where the baud rate equals the channel bandwidth). We measure the minimum ${\widetilde {\mathrm{SNR}}}$ (worst-case SNR) achieved by our method versus {three baselines: a bottleneck Lin–Kernighan heuristic (BLKH \cite{HelsgaunBTSP}), the exact refinement method: Larusic-Punnen \cite{Larusic2012}, and the 2-max bound method (COBTSP \cite{bhar2019channel})}.  

\subsection{Results}

\subsubsection{ {$\widetilde {\mathrm{SNR}}$} vs. Power}
\begin{figure}[!b] 
    \centering
    \subfloat[$n=10$]{
        \includegraphics[width=\linewidth]{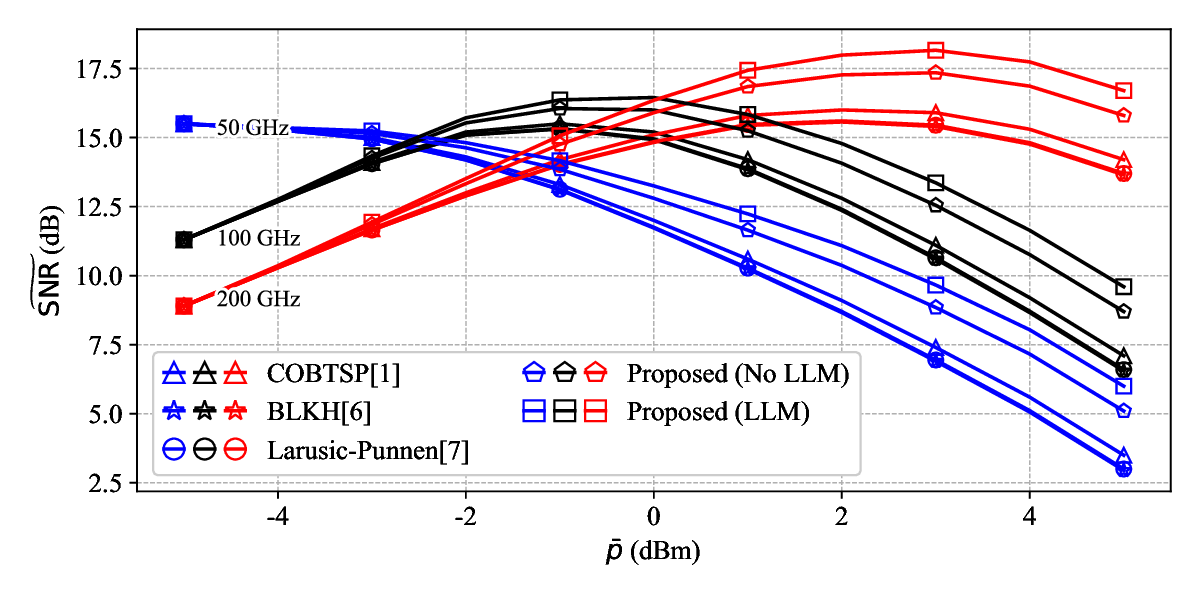}
        \label{fig2a}
    }
    \hfill
    \subfloat[$n=30$]{
        \includegraphics[width=\linewidth]{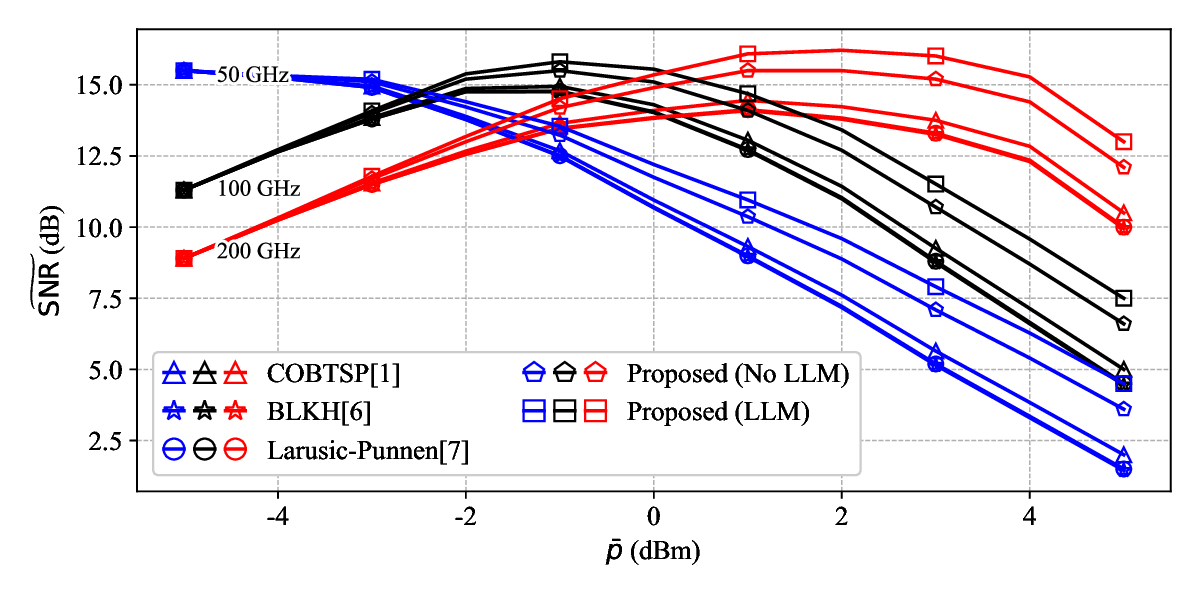}
        \label{fig2b}
    }
    \caption{{$\widetilde {\mathrm{SNR}}$} v.s. power  {(PM-QPSK modulation, $\Delta\left(C_{\pi_i}\right)=F$, $p\left(C_{\pi_i}\right) \in[\bar{p}-5 \mathrm{~dB}, \bar{p}+5 \mathrm{~dB}]$,  for $\forall C_{\pi_i} \in \mathcal{C}$).}}
    \label{fig2}
\end{figure}

{In the low-power (ASE-dominated) regime, $\mathrm{SNR}({C_{\pi_i}}) \approx G({C_{\pi_i}})/G^{\mathrm{ASE}} \propto p({C_{\pi_i}})$, so the curves nearly coincide (the $\leq -4$ dBm region in Fig.~\ref{fig2a} and Fig.~\ref{fig2b}). As the power increases, the system enters the {nonlinearity-limited} regime, where the EGN-predicted NLI scales as $\propto G({C_{\pi_i}})\left(\Phi_{\pi_i} G({C_{\pi_i}})^2+\sum_{j\ne i}\Phi_{{\pi_i},{\pi_j}}G({C_{\pi_j}})^2\right)\sim p({C_{\pi_i}})^3$, leading to $\mathrm{SNR}(C_{\pi_i})\propto p({C_{\pi_i}})^{-2}$ and thus an intermediate optimal launch power $\bar p^*$. A larger channel spacing $F$ substantially suppresses XCI, thereby shifting the optimal power $\bar p$ upward and raising the peak $\widetilde{\mathrm{SNR}}$: experimentally, $200\mathrm{GHz}>100\mathrm{GHz}>50\mathrm{GHz}$. Across all three spacings, the proposed method outperforms the baselines (COBTSP, BLKH, Larusic-Punnen) in the medium-to-high power range. The LLM-enabled variant (Proposed (LLM)) delivers an additional $0.5$-$1$dB improvement over the non-LLM version at most power points. Increasing the channel count from $n=10$ to $n=30$ lowers $\widetilde{\mathrm{SNR}}$ overall by about $1$-$2$ dB and makes the gaps between algorithms more visible, indicating that ordering becomes more valuable in larger systems. }

\subsubsection{ ${\widetilde {\mathrm{SNR}}}$ vs. Number of Channels}
\begin{figure}[h]
\centering
\includegraphics[width=\linewidth]{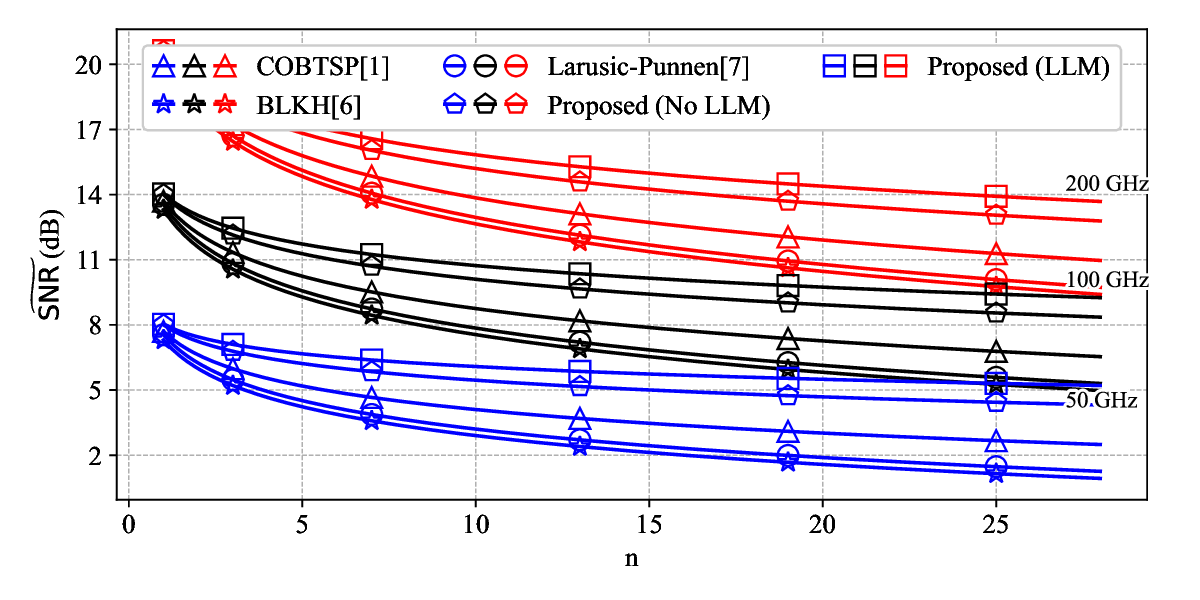}
\caption{{$\widetilde {\mathrm{SNR}}$} vs. number of WDM channels  {(PM-QPSK modulation, $p(C_{\pi_i})=\bar{p}$+5  dBm, $\Delta\left(C_{\pi_i}\right) = F$, for $\forall C_{\pi_i} \in \mathcal{C}$)}}
\label{fig3}
\end{figure}

Next, we examine how the {$\widetilde {\mathrm{SNR}}$} scales with the number of WDM channels $n$ for three spacings ($F\in\{50,100,200\}$ GHz). {As shown in Fig. \ref{fig3}: } $\widetilde{\mathrm{SNR}}$ decreases monotonically with $n$ (a faster drop for small $n$, then a gentle slope). For every spacing, {Proposed (LLM) attains the highest $\widetilde{\mathrm{SNR}}$ over the entire range of $n$; Proposed (No LLM) is consistently second, and the LLM seeding provides an additional $0.4-0.8$ dB over the non-LLM variant. Relative to the baselines (COBTSP/BLKH/Larusic–Punnen), the gain of Proposed (LLM) is typically about $3$ dB, with a mild tendency to increase as $n$ grows.} Larger spacings yield higher $\widetilde{\mathrm{SNR}}$ throughout—200 GHz $>$ 100 GHz $>$ 50 GHz.

\subsubsection{ {$\widetilde {\mathrm{SNR}}$ vs. Bandwidth/Spacing Ratio}}
\begin{figure}[h]
\centering
\includegraphics[width=\linewidth]{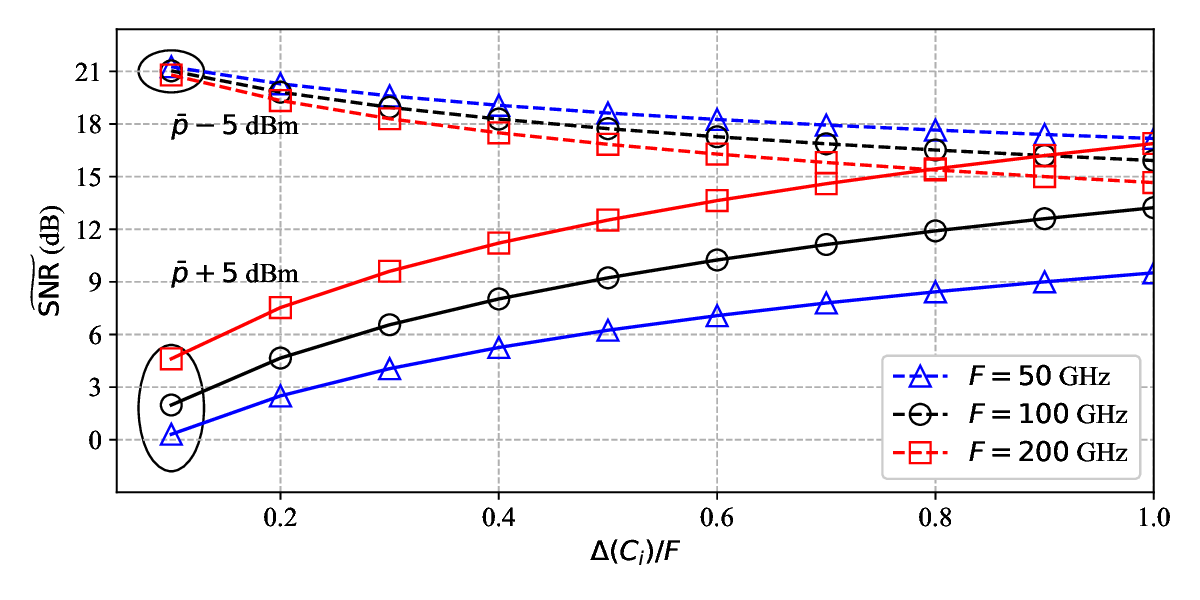}
\caption{{$\widetilde {\mathrm{SNR}}$} vs.  channel bandwidth/spacing ratio at high vs. low launch powers{(PM-QPSK modulation, $n=6$,  $F=200$ GHz). }}
\label{fig4}
\end{figure}

Fig. \ref{fig4} plots the {$\widetilde {\mathrm{SNR}}$} as a function of the channel bandwidth-to-spacing ratio $\Delta/F$, under two launch-power settings (solid lines for high power $\bar p+5$ dBm; dashed lines for low power $\bar p-5$ dBm). {At low launch power (dashed curves), the link is ASE-limited, so increasing the bandwidth-to-spacing ratio $\Delta/F$ lowers each channel’s PSD and SNR declines gently, with little sensitivity to spacing or ordering—hence the tight clustering. At higher launch power (solid curves), the regime is NLI-dominated; here enlarging $\Delta/F$ reduces PSD and suppresses cross/self-channel nonlinear interference faster than the signal, so SNR rises markedly. Grid spacing now matters: larger $F$ weakens the XCI logarithmic kernel, shifting the solid curves upward (most visibly for $F=200$ GHz). Because we optimize the worst-channel SNR, these trends have direct operational meaning: in the ASE regime, ordering and spacing offer minimal leverage and spectral-efficiency considerations should dominate, whereas in the NLI regime, pushing toward larger $\Delta/F$ and, where feasible, larger $F$ lifts the SNR by protecting the most vulnerable (“victim”) channel, making channel ordering most impactful.}

\subsubsection{ {$\widetilde {\mathrm{SNR}}$ vs. XCI Orders}}
\begin{figure}[h]
\centering
\includegraphics[width=\linewidth]{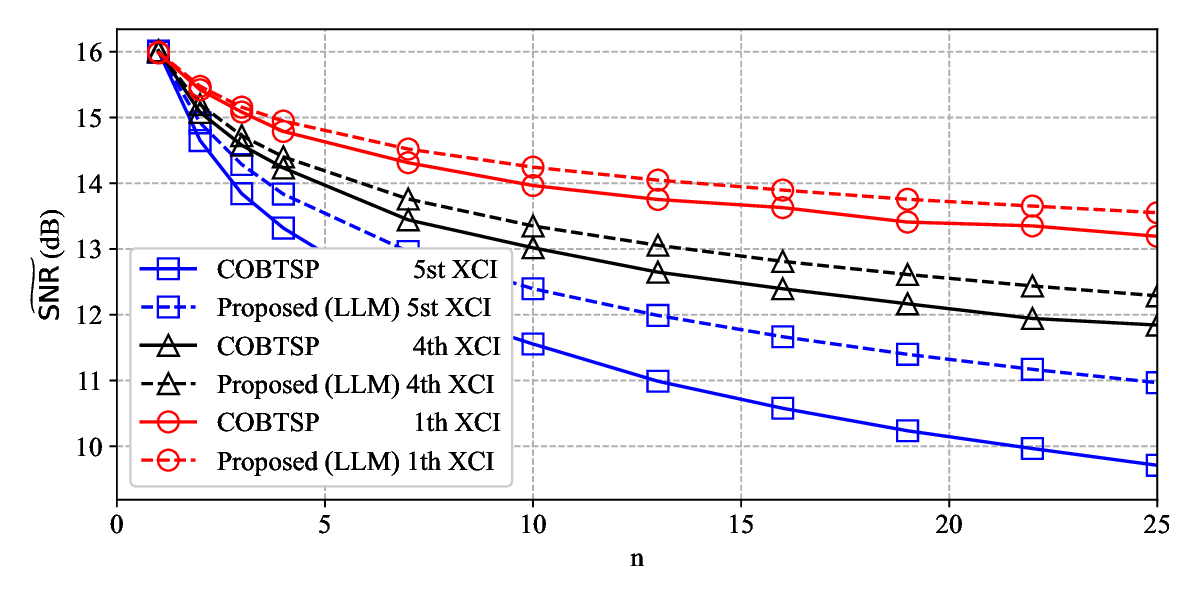}
\caption{{$\widetilde {\mathrm{SNR}}$ vs. XCI Orders {(PM-QPSK modulation,  $p(C_{\pi_i})=\bar{p}=0$  dBm),  $F=\Delta\left(C_{\pi_i}\right)=100$ GHz,  for $\forall C_{\pi_i} \in \mathcal{C}$ }}.}
\label{fig5}
\end{figure}

To test the validity of this assumption, we simulate a worst-case dense scenario ($F=\Delta=50$ GHz) and compare results using different XCI models. Fig. \ref{fig5} shows {$\widetilde {\mathrm{SNR}}$} vs. number of channels for three modeling approaches: (i) 1st-order XCI, which includes only nearest neighbors ($C_{i\pm1}$) in computing NSR as in Eq. \eqref{eq:NSR}; (ii) 4th-order XCI, which includes up to four neighbors on each side ($C_{i\pm1,\pm2,\pm3,\pm4}$); and (iii) 5th-order XCI, which includes up to five neighbors on each side ($C_{i\pm1,\dots,\pm5}$).  As expected, including more interfering neighbors lowers {$\widetilde {\mathrm{SNR}}$} for a given $n$ (since more noise is accounted for), but critically, {the proposed (LLM) algorithm consistently outperforms the representative baseline (CoBTSP) in all cases}. {To maintain clarity, we only show the proposed (LLM) and CoBTSP algorithms in Fig. \ref{fig5}; including all baselines would result in excessive visual clutter}. This demonstrates that our graph-based algorithm, which explicitly optimizes the ordering for adjacent-channel interference, is indeed capturing the dominant effects.

\subsubsection{{Scalability}}

\begin{figure}[h]
\centering
\includegraphics[width=\linewidth]{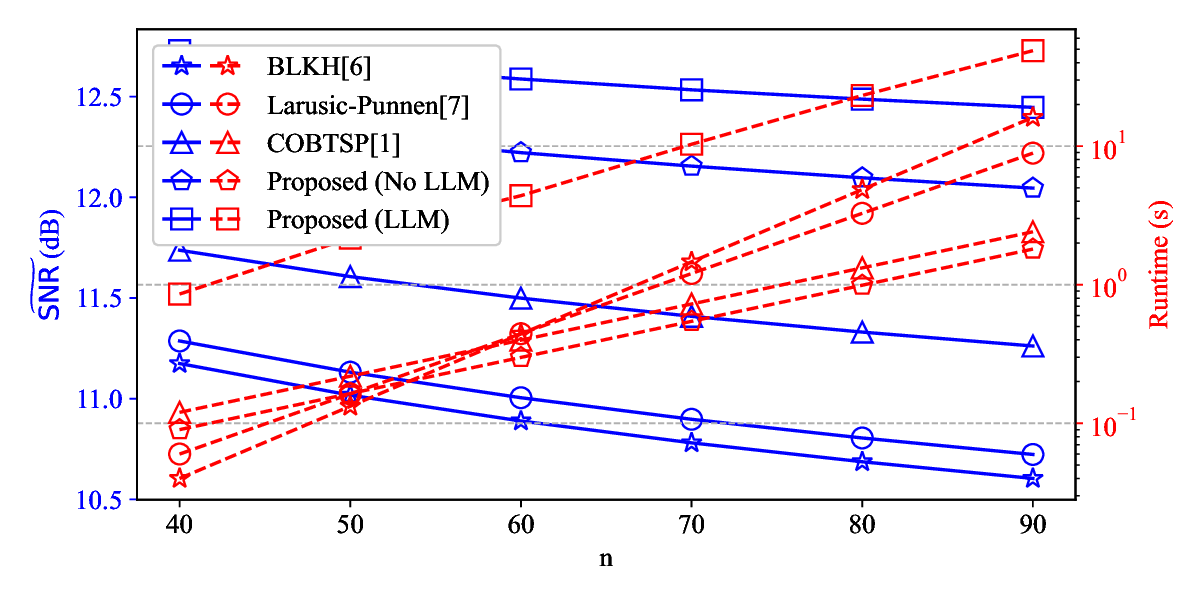}
\caption{$\widetilde {\mathrm{SNR}}$ (blue) and runtime (red) vs. number of channels  {(PM-QPSK modulation, $p(C_{\pi_i})=\bar{p}$+5  dBm, $\Delta\left(C_{\pi_i}\right) = F = 50$ GHz, for $\forall C_{\pi_i} \in \mathcal{C}$)}. }
\label{fig6}
\end{figure}

Finally, Fig. \ref{fig6} presents scalability test. As the number of channels increases from $n=40$ to $n=90$. The $\widetilde{\mathrm{SNR}}$ decreases slightly with $n$ for all methods, yet Proposed (LLM) remains the best over the entire range, with Proposed (No LLM) consistently second. The LLM seeding yields an additional $0.3-0.5$ dB over the non-LLM variant, and a margin of about $0.8-1.2$ dB over the strongest baseline across $n$.  On runtime, all curves exhibits {near} polynomial growth with $n$ ( {Fig. \ref{fig6}, secondary axis, log scale}). COBTSP is the fastest (sub-second at small $n$, a few seconds at $n=90$), while BLKH and Larusic–Punnen increase to the $10^0-10^1$ s range at large $n$. Our sampler without LLM falls in this middle band and remains practical up to $n=90$. The LLM-guided version incurs the highest runtime—roughly $2-3$ times the non-LLM variant, reaching the tens-of-seconds regime at $n=90$—because its cost is dominated by the external LLM  API latency rather than the algorithm complexity.

\subsubsection{{Impact of Modulation Format}}
\begin{figure}[h]
\centering
\includegraphics[width=\linewidth]{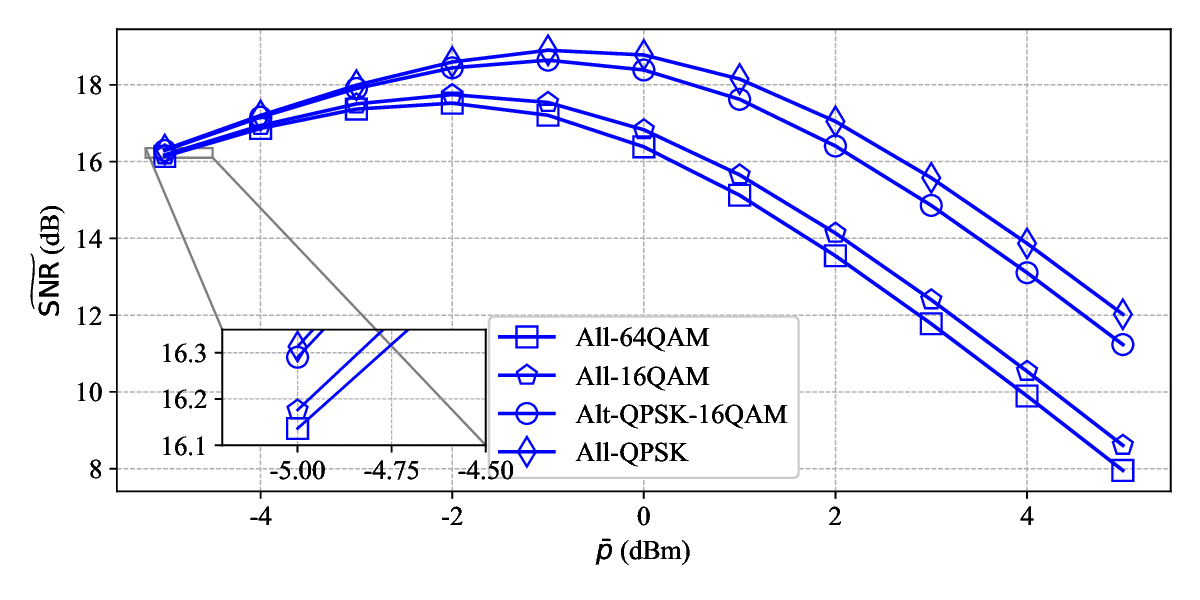}
\caption{{$\widetilde {\mathrm{SNR}}$ vs. average power, with different modulation format ($n=6$, $F=\Delta(C_{\pi_i})$=100 GHz,  for $\forall C_{\pi_i} \in \mathcal{C}$).}}
\label{fig7}
\end{figure}

{We use PM-constellation EGN factors $\Phi_{\mathrm{QPSK}}=1,\ \Phi_{\mathrm{16QAM}}=\tfrac{17}{25}\approx0.68,\Phi_{\mathrm{64QAM}}=\tfrac{13}{21}\approx0.62$; Alt-QPSK-16QAM: even channels $\Phi=1.0$, odd channels $\Phi=0.68$. As shown in Fig. \ref{fig7}, higher-order formats (16/64QAM) consistently achieve {lower} $\widetilde{\mathrm{SNR}}$ than QPSK, with mixed constellations in between. The separation is visible at low powers (-5 dBm) and widens with increasing launch power, since lower $\Phi$ values {incur} XCI growth.}

\section{{Complexity Analysis of The Proposed BTSP Solver}}
{Each channel maintains candidate set of at most $O(n)$ successors are considered at one decision point.  In probabilistic sampling phase, $3n^2$ permutations of length $n$ are sampled, incurring $O(n^3)$ edge evaluations. In refinement phase, $3n^2$ permutations refined for $O(\sqrt n)$ iterations, for an additional $O(n^{2.5})$ edge evaluations. Hence the cost is dominated by the sampling phase and scales as $O(n^3)$, which is polynomial.}

\section{Conclusion}
We address the channel ordering problem in EONs, which directly impacts adjacent-channel interference and overall signal quality, by formulating this challenge as a BTSP, we propose a scalable solver that enhances SNR fairness. Experiments show a clear split: at low launch power the link is ASE-limited (ordering/spacing matter little), while at higher power it is nonlinearity-limited, where larger spacing and a higher bandwidth-to-spacing ratio lift the worst-channel SNR. As channels increase, worst-channel SNR drops and ordering becomes more valuable. The proposed method—especially the LLM-seeded variant—consistently leads (0.4–1.0 dB over non-LLM, 3 dB over baselines) with practical runtimes up to 90 channels, and higher-order modulations yield higher worst-channel SNR, with the gap widening at higher powers.

\bibliographystyle{IEEEtran}
\bibliography{references.bib}

@article{Larusic2012,
  author    = {Jozo J. Larušić and Abraham P. Punnen},
  title     = {A Linear Time Algorithm for the Bottleneck Traveling Salesman Problem},
  journal   = {Computers \& Operations Research},
  volume    = {39},
  number    = {7},
  year      = {2012},
  pages     = {1605--1611},
  doi       = {10.1016/j.cor.2011.08.008}
}

@misc{HelsgaunBTSP,
  author       = {Keld Helsgaun},
  title        = {{Solving the Bottleneck Traveling Salesman Problem with LKH}},
  year         = {2016},
  howpublished = {\url{http://webhotel4.ruc.dk/~keld/research/LKH/BLKH-report.pdf}},
  note         = {Accessed: 2025-04-25}
}

@article{fang2024reconfigurable,
  title={Reconfigurable intelligent surface assisted free space optical information and power transfer},
  author={Fang, Wen and Chen, Wen and Wu, Qingqing and Wang, Kunlun and Zhang, Shunqing and Liu, Qingwen and Li, Jun},
  journal={IEEE Internet of Things Journal},
  volume={11},
  number={18},
  pages={30260--30277},
  year={2024},
  publisher={IEEE}
}

@article{yang2023large,
  title={Large language models as optimizers},
  author={Yang, Chengrun and Wang, Xuezhi and Lu, Yifeng and Liu, Hanxiao and Le, Quoc V and Zhou, Denny and Chen, Xinyun},
  journal={arXiv preprint arXiv:2309.03409},
  year={2023}
}

@article{romera2024mathematical,
  title={Mathematical discoveries from program search with large language models},
  author={Romera-Paredes, Bernardino and Barekatain, Mohammadamin and Novikov, Alexander and Balog, Matej and Kumar, M Pawan and Dupont, Emilien and Ruiz, Francisco JR and Ellenberg, Jordan S and Wang, Pengming and Fawzi, Omar and others},
  journal={Nature},
  volume={625},
  number={7995},
  pages={468--475},
  year={2024},
  publisher={Nature Publishing Group UK London}
}

@article{liu2024evolution,
  title={Evolution of heuristics: Towards efficient automatic algorithm design using large language model},
  author={Liu, Fei and Tong, Xialiang and Yuan, Mingxuan and Lin, Xi and Luo, Fu and Wang, Zhenkun and Lu, Zhichao and Zhang, Qingfu},
  journal={arXiv preprint arXiv:2401.02051},
  year={2024}
}

@article{ye2024reevo,
  title={Reevo: Large language models as hyper-heuristics with reflective evolution},
  author={Ye, Haoran and Wang, Jiarui and Cao, Zhiguang and Berto, Federico and Hua, Chuanbo and Kim, Haeyeon and Park, Jinkyoo and Song, Guojie},
  journal={arXiv preprint arXiv:2402.01145},
  year={2024}
}

@article{bhar2019channel,
  title={Channel allocation in elastic optical networks using traveling salesman problem algorithms},
  author={Bhar, Chayan and Agrell, Erik and Keykhosravi, Kamran and Karlsson, Magnus and Andrekson, Peter A},
  journal={Journal of Optical Communications and Networking},
  volume={11},
  number={10},
  pages={C58--C66},
  year={2019},
  publisher={OSA}
}

@article{Poggiolini,
author = {P. Poggiolini and Y. Jiang},
journal = {J. Lightwave Technol.},
number = {3},
pages = {458--480},
publisher = {Optica Publishing Group},
title = {Recent Advances in the Modeling of the Impact of Nonlinear Fiber Propagation Effects on Uncompensated Coherent Transmission Systems},
volume = {35},
month = {Feb},
year = {2017}}

@article{roberts2018measurement,
  title={Measurement-based optimization of channel powers with non-Gaussian nonlinear interference noise},
  author={Roberts, Ian and Kahn, Joseph M},
  journal={Journal of Lightwave Technology},
  volume={36},
  number={13},
  pages={2746--2756},
  year={2018},
  publisher={OSA}
}

@article{roberts2016convex,
  title={Convex channel power optimization in nonlinear WDM systems using Gaussian noise model},
  author={Roberts, Ian and Kahn, Joseph M and Boertjes, David},
  journal={Journal of Lightwave Technology},
  volume={34},
  number={13},
  pages={3212--3222},
  year={2016},
  publisher={IEEE}
}

@book{applegate2006traveling,
  title={The traveling salesman problem: a computational study},
  author={Applegate, David L},
  volume={17},
  year={2006},
  publisher={Princeton university press}
}

@article{Qin1994Bellman,
  title={Bellman optimality principle(I)},
  author={Yuyuan Qin},
  journal={MATHEMATICA APPLICATA},
  volume={7},
  number={3},
  pages={349-354},
  year={1994},
}

@article{kinable2017exact,
  title={Exact algorithms for the equitable traveling salesman problem},
  author={Kinable, Joris and Smeulders, Bart and Delcour, Eline and Spieksma, Frits CR},
  journal={European Journal of Operational Research},
  volume={261},
  number={2},
  pages={475--485},
  year={2017},
  publisher={Elsevier}
}

@article{helsgaun2000effective,
  title={An effective implementation of the Lin--Kernighan traveling salesman heuristic},
  author={Helsgaun, Keld},
  journal={European journal of operational research},
  volume={126},
  number={1},
  pages={106--130},
  year={2000},
  publisher={Elsevier}
}

\end{document}